\documentstyle[amsfonts,amsmath,amssymb,epsfig,cite,a4wide]{article}
\newcommand{\vb}[1]{{\mathbf{#1}}}
\newcommand{\lb}[1]{\label{#1}}

\newcommand{\bc}{\begin{center}}
\newcommand{\ec}{\end{center}}
\newcommand{\be}{\begin{equation}}
\newcommand{\ee}{\end{equation}}
\newcommand{\bea}{\begin{eqnarray}}
\newcommand{\eea}{\end{eqnarray}}
\newcommand{\ba}[1]{\begin{array}{#1}}
\newcommand{\ea}{\end{array}}

\newcommand{\bt}[1]{\begin{table}[ht]\centering\begin{tabular}{#1}}
\newcommand{\et}[1]{\end{tabular}\caption{\small#1}\end{table}}

\begin{document}

\thispagestyle{empty}

\begin{flushright} 
{\tt hep-th/0510063}\\
{\sl May 2006}
\end{flushright}

\begin{center}

\vspace{1.5 truecm}

{\large\bf{Explicit Actions for Electromagnetism with Two Gauge Fields\\ with Only one Electric and one Magnetic Physical Fields}}
\vspace{1.5 truecm}\\

{\small{\it
CENTRA, Instituto Superior T\'ecnico, Av. Rovisco Pais, 1049-001 Lisboa, Portugal\\Dpt. F\'{\i}s., Univ. Beira Int., Rua Marqu\^es D'\'{A}vila e Bolama, 6200-081 Covilh\~a, Portugal}}

\vfill
{\bf\sc Abstract}\\[5mm]

\begin{minipage}{15cm}
We extend the work of Mello et al. based in Cabbibo and Ferrari
concerning the description of electromagnetism with two gauge fields
from a variational principle, i.e. an action. We provide a systematic
independent derivation of the allowed actions which have only one
magnetic and one electric physical fields and are invariant under
the discrete symmetries $P$ and $T$. We conclude that neither
the Lagrangian, nor the Hamiltonian, are invariant under the electromagnetic
duality rotations. This agrees with the weak-strong coupling
mixing characteristic of the duality due to the Dirac quantization condition
providing a natural way to differentiate dual theories related by the duality
rotations (the energy is not invariant). Also the standard electromagnetic duality
rotations considered in this work violate both $P$ and $T$ by inducing Hopf terms
(theta terms) for each sector and a mixed Maxwell term. The canonical structure of
the theory is briefly addressed and the \textit{magnetic}
gauge sector is interpreted as a ghost sector.
\end{minipage}

\end{center}

\vfill
\begin{flushleft}
\end{flushleft}

\newpage
\pagestyle{empty}

\tableofcontents

\newpage

\pagestyle{plain}
\setcounter{page}{1}

\section{Introduction and Discussion of Results\lb{sec.intro}}

The seminal works of Dirac~\cite{Dirac} introduced the famous charge quantization relation $eg=n$
which is obtained in the presence of both electric and magnetic poles (charges). The existence
of both electric and magnetic charges raised the problem of a variational description of electromagnetism
from an action that could actually contain explicitly both types of charges. Also it is widely
accepted that in order to achieve that goal one must consider a description in terms of gauge fields
which minimally couple to both currents, so necessarily we need to consider the existence of two distinct
gauge fields, $A$ that couples to ordinary electric currents and $C$ that couples to magnetic
current~\cite{CF,Schwinger,Zwa_01,Zwa_02,Zwa_03}. One possible approach first considered by 
Cabbibo and Ferrari~\cite{CF} is to consider two physical gauge fields $A$ and $C$. Although
this approach preserves both time-space isotropy and Lorentz invariance has the drawback of the inexistence
of experimental observable effects of the second gauge field. Another approach have been to consider mechanisms
that starting from a theory with two gauge fields give us only one physical gauge field, either by considering
solutions (constraints) for the second gauge field~\cite{Schwinger,Zwa_01,Zwa_02,Zwa_03} (this approach
has the drawback of not preserving space isotropy or not preserving Lorentz invariance)
or by considering a very massive second gauge field~\cite{Sing_01}.
Yet another very simple approach is to consider electromagnetism as an
effective theory of an extended theory with two gauge fields such that one gauge field is
fixed by the second gauge field obeying the equations of motion~\cite{outro}.

In Mello et al.~\cite{action_00} it is build for the first time an explicit action for electromagnetism
with two gauge fields based in the work of Cabbibo and Ferrari~\cite{CF}. In here we build a similar
lower order action with two gauge fields $A$ and $C$ of the gauge group $U(1)\times U(1)$.
In order to accomplish it we take an independent approach of the original work~\cite{action_00}
by studying in detail and systematically the desired properties of such an action. First we note that due to the
different nature of $A$ and $C$ under the discrete symmetries of parity $P$ and time inversion $T$~\cite{Jackson,Sing_02},
standard electromagnetic duality~\cite{Jackson,olive} violates $P$ and $T$ symmetries. So it is desirable that under an
electromagnetic duality transformation our action gains terms that explicit violate these
symmetries~\footnote{This argument is not completely closed once there are ways of implementing duality rotations
that preserve $P$ and $T$ symmetries~\cite{new_01,Sing_03}.}.
Secondly we demand that there are only one electric and one magnetic physical fields. Implicitly this assumption
means that the group charge flux of each of the $U(1)$'s is of the same nature of the topological flux
of the other $U(1)$ group. The action suggested coincides (up to a sign choice) with the one of~\cite{action_00}
and consists of two Maxwell terms with opposite relative sign, one for each of the gauge fields and a topological
cross Hopf term that mixes both gauge sectors allowing for the desired characteristics,
\bea
S^{\hat{\epsilon}}_{\mathrm{Max}_-}=-\int_M&\displaystyle\left[\frac{\sqrt{-g}}{4e^2}F_{\mu\nu}F^{\mu\nu}-\frac{\sqrt{-g}}{4g^2}G_{\mu\nu}G^{\mu\nu}-\frac{\hat{\epsilon}}{4e\,g}\epsilon^{\mu\nu\rho\delta}F_{\mu\nu}G_{\rho\delta}+\frac{1}{e}(A_\mu-\hat{\epsilon}\tilde{C}_\mu)J_e^\mu-\frac{1}{g}(\hat{\epsilon}C_\mu+\tilde{A}_\mu) J_g^\mu\right]\ ,
\nonumber
\eea
with $\hat{\epsilon}=\pm 1$ corresponding to the two physical fields
\bea
\ba{rcl}
E^i&=&\displaystyle \frac{1}{e}F^{0i}-\frac{\hat{\epsilon}}{2g}\epsilon^{0ijk}G_{jk}\\[5mm]
B^i&=&\displaystyle\frac{\hat{\epsilon}}{g}G^{0i}+\frac{1}{2e}\epsilon^{0ijk}F_{jk}\\[5mm]
\ea
\nonumber
\eea
However the Maxwell terms of each of the gauge sectors have opposite sign,
this has no consequences at classical level but at quantum level allows negative energy solutions which clearly 
violates causality. There are two approaches to overcome this problem. We can consider the $C$ field to be a ghost,
this means that upon quantization it has the opposite spin-statistics relations than the one of standard fields
and therefore it has anti-commutation relations~\cite{PS}, such kind of theories both with a matter and a
ghost sector were introduced in cosmology by Linde~\cite{Linde}. Alternatively we can consider some mechanism
that allows for a classical treatment of the $C$ field, as examples we have in cosmology the Phantom matter
models~\cite{Phantom} and a dynamical symmetry breaking mechanism~\cite{outro} that allows a effective electric description
of the theory. Also compatible with this last mechanism we can give a vacuum-expectation-value to the $C$ field
that renders an effective Proca mass to the standard photon, the $A$ field~\cite{Tito_01,Tito_02}.

\section{Electromagnetic Duality\lb{sec.dual}}

The study of theories with two gauge fields were first considered by Cabbibo and Ferrari~\cite{CF}.
More recently several studies addressed electromagnetic duality with two gauge
fields, namely in~\cite{Sing_02} a explicit electromagnetic duality in terms of the gauge
fields is presented. Here we review these results.

\subsection{The Original Duality}

The generalized Maxwell equations with both Electric and Magnetic currents~\cite{Jackson} read
\be
\ba{rcl}
\nabla.\vb{E}&=&\rho_e\\[5mm]
\nabla.\vb{B}&=&\rho_g\\[5mm]
\dot{\vb{B}}+\nabla\times \vb{E}&=&-\vb{J}_g\\[5mm]
\dot{\vb{E}}-\nabla\times \vb{B}&=&-\vb{J}_e\ .
\ea
\lb{Maxwell}
\ee
This equation obey the well known electromagnetic duality which rotates the electric and magnetic
fields and currents~\cite{olive}
\be
\left\{
\ba{rcl}
\vb{E}&\to&\cos(\theta)\vb{E}+\sin(\theta)\vb{B}\\[5mm]
\vb{B}&\to&-\sin(\theta)\vb{E}+\cos(\theta)\vb{B}\\[5mm]
J_e&\to&\cos(\theta)J_e+\sin(\theta)J_g\\[5mm]
J_g&\to&-\sin(\theta)J_e+\cos(\theta)J_g
\ea
\right.
\lb{duality_EB}
\ee
where $J=(\rho,\vb{J})$ stand for the 4-vector current densities.

\subsection{Duality with two Gauge Fields}

In order to build an action
for electromagnetism with magnetic monopoles it is necessary to consider two $U(1)$ gauge fields which minimally couple
to the external electric and magnetic current densities. By introducing gauge fields one is led to the question
weather the above duality can be extended to a duality of gauge fields instead of the electric and magnetic fields
(i.e. the gauge field connections). By considering that both gauge fields have true physical degrees of freedom it
is possible to elevate the duality to a transformation of those gauge fields as have been shown in~\cite{Sing_02}.
In~\cite{Sing_02} the electric and magnetic fields are defined as
\be
\ba{rcl}
E^i&=&\displaystyle\frac{1}{2e}\,F^{0i}-\frac{1}{4g}\,\epsilon^{ijk}G_{jk}\\[5mm]
B^i&=&\displaystyle\frac{1}{2g}\,G^{0i}+\frac{1}{4e}\,\epsilon^{ijk}F_{jk}\\[5mm]
\ea
\lb{EB}
\ee
where $F=dA$ and $G=dC$ are the gauge connections of the gauge fields $A$ and $C$.
In section~\ref{sec.fields} we will properly discuss the physical field definitions, for
the moment being we use these definitions which can be found in the literature.
The electromagnetic duality reads now
\be
\ba{rcl}
\displaystyle\frac{1}{e}\,F^{0i}-\frac{1}{2g}\,\epsilon^{ijk}G_{jk}&\to&\displaystyle\ \cos(\theta)\left(\frac{1}{e}\,F^{0i}-\frac{1}{2g}\,\epsilon^{ijk}G_{jk}\right)+\sin(\theta)\left(\frac{1}{g}\,G^{0i}+\frac{1}{2e}\,\epsilon^{ijk}F_{jk}\right)\\[6mm]
\displaystyle\frac{1}{g}\,G^{0i}+\frac{1}{2e}\,\epsilon^{ijk}F_{jk}&\to&\displaystyle-\sin(\theta)\left(\frac{1}{e}\,F^{0i}-\frac{1}{2g}\,\epsilon^{ijk}G_{jk}\right)+\cos(\theta)\left(\frac{1}{g}\,G^{0i}+\frac{1}{2e}\,\epsilon^{ijk}F_{jk}\right)
\ea
\ee
There are two ways to implement these transformations, either in terms of each $U(1)$ gauge sectors independently
or mixing both gauge sectors. If we consider each sector independently we obtain the
standard electromagnetic transformations for each of the connections $F$ and $G$
\be
\ba{rcl}
F^{0i}&\to&\cos(\theta)\,F^{0i}+\sin(\theta)\,\frac{1}{2}\epsilon^{ijk}F_{jk}\\[5mm]
F_{jk}&\to&\sin(\theta)\,\frac{1}{2}\epsilon_{ijk}F^{0i}+\cos(\theta)\,F_{jk}\\[5mm]
G^{0i}&\to&\cos(\theta)\,G^{0i}+\sin(\theta)\,\frac{1}{2}\epsilon^{ijk}G_{jk}\\[5mm]
G_{jk}&\to&\sin(\theta)\,\frac{1}{2}\,\epsilon^{ijk}G^{0i}+\cos(\theta)\,G_{jk}
\ea
\lb{duality_independent}
\ee
These transformations are not compatible with a transformation of the gauge fields because
the $(0i)$ components transform differently from the components $(ij)$.

If we consider mixing between both sectors we can rewrite the electromagnetic
duality in terms of the gauge fields or respective connections~\cite{Sing_02}
\be
\left\{
\ba{rcl}
F&\to&\displaystyle\ \cos(\theta)\,F+\sin(\theta)\,\frac{e}{g}G\\[5mm]
G&\to&\displaystyle-\sin(\theta)\,\frac{g}{e}F+\cos(\theta)\,G\\[5mm]
A&\to&\displaystyle\ \cos(\theta)\,A+\sin(\theta)\,\frac{e}{g}C\\[5mm]
C&\to&\displaystyle-\sin(\theta)\,\frac{g}{e}A+\cos(\theta)\,C
\ea
\right.
\lb{duality}
\ee
There is a very simple argument to choose the
second kind of duality~(\ref{duality}) and exclude the possibility of the transformations~(\ref{duality_independent}).
Let us consider the Lorentz gauge (or Lorentz condition) for both gauge fields
$\partial_\mu A^\mu=\partial_\mu C^\mu=0$ and assume regular gauge fields
(meaning without discontinuities) such that the Bianchi identities are obeyed
$\epsilon^{\mu\nu\rho\delta}\partial_\nu\partial_\rho A_\delta=\epsilon^{\mu\nu\rho\delta}\partial_\nu\partial_\rho C_\delta=0$.
Then the Maxwell equations~(\ref{Maxwell}) read simply~\cite{Sing_02}
\be
\ba{rcl}
\Delta A^\mu&=&J^\mu_e\\[5mm]
\Delta C^\mu&=&J^\mu_g
\ea
\lb{Maxwell_decoupled}
\ee
where the Laplacian is $\Delta=\partial_\mu\partial^\mu$. Taking in account the duality transformations
for the current densities  expressed in~(\ref{duality_EB}) we conclude straight away that
only~(\ref{duality}) correctly transform the Maxwell equations for these
particular \textit{standard} conditions. Here particular means that the gauge choice is not unique, we
could have some other gauge fixing prescription and generally we can have discontinuities on the
gauge fields such that the Bianchi identity is not obeyed everywhere. As a example there are the cases
of the Dirac string~\cite{Dirac} or equivalently the nontrivial fiber-bundle of Wu and Yang~\cite{YW}.
However regular gauge fields describe most of physical applications and must therefore be a possible choice.
There is however a serious problem concerning these equations, the two $U(1)$ gauge fields
are completely decoupled and we obtain two different interactions corresponding to each of the gauge fields instead
of only one as in standard electromagnetism. Our main aim in the remaining of this work is how to
obtain one only interaction described by two physical gauge fields.

So we have reviewed how to elevate electromagnetic duality of the Maxwell equations in terms of
the electric and magnetic fields to a electromagnetic duality in terms of the gauge fields.
Next we will briefly describe how the discrete symmetries act on the several fields
and how electromagnetic duality breaks parity and time inversion.

\subsection{Discrete Symmetries: $P$ and $T$ Violation}

We proceed to resume the known results for parity $P$ and time inversion $T$
for the electromagnetic physical quantities. The remaining discrete symmetry is Charge Conjugation $C$
and plays no role in the following discussion.

Parity ($P$) stands for the inversion of spatial coordinates and time-inversion ($T$) stands for the inversion of the time coordinate.
Under these discrete symmetries the fields and current densities transform as~\cite{Jackson}
\be
\ba{rrcl}
P:\ &x^i&\to &-x^i\\[5mm]
    &E^i&\to &-E^i\\[5mm]
    &B^i&\to &+B^i\\[5mm]
    &\rho_e&\to &+\rho_e\\[5mm] 
    &J^i_e&\to &-J^i_e\\[5mm] 
    &\rho_g&\to &-\rho_g\\[5mm]
    &J^i_g&\to &+J_g
\ea\ \ \ \ \ 
\ba{rrcl}
T:\ &t&\to &-t\\[5mm]
    &E^i&\to &+E^i\\[5mm]
    &B^i&\to &-B^i\\[5mm]
    &\rho_e&\to &+\rho_e\\[5mm] 
    &J^i_e&\to &-J^i_e\\[5mm] 
    &\rho_g&\to &-\rho_g\\[5mm]
    &J^i_g&\to &+J_g
\ea
\ee

Electric and magnetic fields transform differently under $P$ and $T$
being respectively vectors and pseudo-vectors. Accordingly also the electric and magnetic
currents have the same properties~\cite{Jackson}. Then necessarily the gauge fields $A$ and $C$ also
have to transform accordingly as vectors and pseudo-vectors~\cite{Sing_02}. The
most straight forward way to show this is by considering an action for electromagnetism
such that the electric and magnetic current densities are minimally coupled
to the gauge fields $A$ and $C$ respectively (we will return to this discussion later).
Demanding invariance of the action under $P$ and $T$ imposes the gauge field $C$
to transform as a pseudo-vector. We note that the field definitions~(\ref{EB}) agree with
these results. Then for the two gauge fields and respective gauge connections
we have the discrete transformations
\be
\ba{rrcl}
P:\ &A^0&\to &+A^0\\[5mm]
    &A^i&\to &-A^i\\[5mm]
    &C^0&\to &-C^0\\[5mm]
    &C^i&\to &+C^i\\[5mm]
    &F^{0i}&\to &-F^{0i}\\[5mm]
    &F^{ij}&\to &+F^{ij}\\[5mm] 
    &G^{0i}&\to &+G^{0i}\\[5mm]
    &G^{ij}&\to &-G^{ij}
\ea\ \ \ \ 
\ba{rrcl}
T:\ &A^0&\to &+A^0\\[5mm]
    &A^i&\to &-A^i\\[5mm]
    &C^0&\to &-C^0\\[5mm]
    &C^i&\to &+C^i\\[5mm]
    &F^{0i}&\to &+F^{0i}\\[5mm]
    &F^{ij}&\to &-F^{ij}\\[5mm] 
    &G^{0i}&\to &-G^{0i}\\[5mm]
    &G^{ij}&\to &+G^{ij}
\ea\ \ \ \ 
\lb{CPT}
\ee

We want now to show that neither $P$ nor $T$ are maintained by the standard duality rotations~(\ref{duality_EB})
or equivalently~(\ref{duality}). Here we consider duality as a global transformation independent of space-time
coordinates such that the angle $\theta$ is an exterior parameter to the theory
used in the redefinition of the fields. Therefore it does not depend on the space-time coordinates
and transform as a scalar with respect to the discrete symmetries $P$ and $T$.

We can see explicitly that the duality transformations mix vector with pseudo-vectors such that
\be
\ba{rrcl}
P:\ &\vb{\tilde{E}}&=   &\cos(\theta)\vb{E}+\sin(\theta)\vb{B}\\[5mm]
    &              &\to &-\cos(\theta)\vb{E}+\sin(\theta)\vb{B}\\[7mm]
    &\vb{\tilde{B}}&=   &-\sin(\theta)\vb{E}+\cos(\theta)\vb{B}\\[5mm]
    &              &\to &\sin(\theta)\vb{E}+\cos(\theta)\vb{B}\ .
\ea
\lb{dualEB_P}
\ee
Clearly $\vb{\tilde{E}}$ and $\vb{\tilde{B}}$ are not transformed to $-\vb{\tilde{E}}$ and
$\vb{\tilde{B}}$ under parity as they should. The same argument follows for $T$
\be
\ba{rrcl}
T:\ &\vb{\tilde{E}}&=   &\cos(\theta)\vb{E}+\sin(\theta)\vb{B}\\[5mm]
    &              &\to &\cos(\theta)\vb{E}-\sin(\theta)\vb{B}\\[7mm]
    &\vb{\tilde{B}}&=   &-\sin(\theta)\vb{E}+\cos(\theta)\vb{B}\\[5mm]
    &              &\to &-\sin(\theta)\vb{E}-\cos(\theta)\vb{B}
\ea
\lb{dualEB_T}
\ee
and the redefined fields do not transform correctly under $T$.
The current duality transformations~(\ref{duality_EB}) behaves in the same way.

Charge conjugation $C$ is not a space-time symmetry, it exchanges
particles with antiparticles. At classical level this is simply equivalent to change
the sign of the current densities and it is preserved by electromagnetic duality.

So, to summarize, at the level of single fields, the electromagnetic duality
preserves $C$, $PT$ and $CPT$ while it violates $P$, $T$, $CP$ and $CT$.

The issue of $P$ and $T$ violation by the existence of dyons with both electric and magnetic charge can
be found in~\cite{Jackson}. As for $P$ and $T$ violation by electromagnetic duality is discussed in~\cite{Sing_03}.
The argument is generic and applicable to the original duality transformations~(\ref{duality_EB})
independently of considering a gauge field description of electromagnetism. Also we point out that upon
redefinition of the fields one may as well redefine $P$ and $T$, but in order to do so
one would be changing the space-time interpretation of the discrete symmetries and necessarily
redefining the action of the Lorentz group. This could be interpreted then as an extended duality
of space-time. An alternative interesting construction is to consider $\theta$ to be a pseudo-scalar~\cite{Sing_03},
in this way we manage to obtain a duality that preserves the discrete symmetries. Also it is possible
to gauge the duality by considering $\theta=\theta(x)$ to be an additional gauge
parameter~\cite{new_01,Sing_04} (in this works the duality rotations constitute one further distinct $U(1)$ group).

Here we are considering $\theta$ to be a parameter exterior to the theory that transforms as a scalar,
then, although the discrete symmetries violations are not explicit in the equations of motion,
at the level of the action (a Lagrangian formulation of the theory) they will be explicit.
As we will see in detail electromagnetic duality induces $P$ and $T$ violating terms.

In addition we will demand that there is only one electric and one magnetic
physical gauge fields. This requirement is going to reduce the allowed actions.

\section{Gauge Sector\lb{sec.gauge}}

In this section we will build a $U(1)\times U(1)$ gauge action such that the physical
electric and magnetic fields are identified with the definitions~(\ref{EB}).
In order to do so one expects that the group charge flux of each of the $U(1)$'s
is coupled to the topological charge flux of the other $U(1)$. It is also desirable that
a classical description of electromagnetism preserves both parity~$P$ and time inversion~$T$
(see for instance~\cite{Jackson} for a discussion on this topic).
So we are further demanding our action to be invariant under these discrete symmetries.
In addition, and from the discussion on the last section, we expect
that under an electromagnetic rotation our action explicitly gains terms that
violate $P$ and $T$. This will be the case.

\subsection{Possible Actions}

Let us consider all the possible lower order terms that are Lorentz and gauge invariant.
First we list the lower order terms containing the gauge connections $F$ and $G$ which are invariant
under $P$ and $T$
\be
\ba{rcl}
{\mathcal{L}}_{\mathrm{Maxwell}_{FF}}&=&\displaystyle-\frac{1}{4\,e^2}F_{\mu\nu}F^{\mu\nu}\\[5mm]
{\mathcal{L}}_{\mathrm{Maxwell}_{GG}}&=&\displaystyle-\frac{1}{4\,g^2}G_{\mu\nu}G^{\mu\nu}\\[5mm]
{\mathcal{L}}_{\mathrm{Hopf}_{FG}}&=&\displaystyle-\frac{1}{4\,eg}\epsilon^{\mu\nu\rho\delta} F_{\mu\nu}G_{\rho\delta}\ .
\ea
\lb{L_inv}
\ee
The last term is a cross Hopf term (or theta term). To show that it is invariant let us rewrite
the expression as ${\mathcal{L}}_{\mathrm{Hopf}}=2\epsilon^{0ijk}(F_{0i}G_{jk}+G_{0i}F_{jk})$, then
we see from~(\ref{CPT}) that $F_{0i}$ and $G_{0i}$ always transform in the same way as $G_{jk}$ and
$F_{jk}$ (respectively) such that ${\mathcal{L}}_{\mathrm{Hopf}_{FG}}$ is invariant under any of
the discrete symmetries $P$ and $T$.

The remaining possible lower order terms which are Lorentz and gauge invariant are not invariant under $P$ and $T$.
They are the cross Maxwell term and the usual Hopf (or theta) terms for each of the gauge sectors
\be
\ba{rcl}
{\mathcal{L}}_{\mathrm{Maxwell}_{FG}}&=&\displaystyle-\frac{1}{4\,eg}F_{\mu\nu}G^{\mu\nu}\\[5mm]
{\mathcal{L}}_{\mathrm{Hopf}_{FF}}&=&\displaystyle-\frac{1}{4\,e^2}\epsilon^{\mu\nu\rho\delta} F_{\mu\nu}F_{\rho\delta}\\[5mm]
{\mathcal{L}}_{\mathrm{Hopf}_{GG}}&=&\displaystyle-\frac{1}{4\,g^2}\epsilon^{\mu\nu\rho\delta} G_{\mu\nu}G_{\rho\delta}
\ea
\lb{L_var}
\ee
To show that they are not invariant under $P$ and $T$ we note that, from equation~(\ref{CPT}),
$(F_{0i},F_{ij})$ and $(G_{0i},G_{ij})$ transform in the opposite way under $P$ and $T$ such that
the cross Maxwell term transforms as ${\mathcal{L}}_{\mathrm{Maxwell}_{FG}}\to-{\mathcal{L}}_{\mathrm{Maxwell}_{FG}}$.
Concerning the Hopf terms we note that $F_{0i}$ and $G_{0i}$ transform in the opposite way than
$F_{ij}$ and $G_{ij}$ (respectively) under $P$ and $T$ such that
${\mathcal{L}}_{\mathrm{Hopf}_{FF}}\to-{\mathcal{L}}_{\mathrm{Hopf}_{FF}}$ and
${\mathcal{L}}_{\mathrm{Hopf}_{GG}}\to-{\mathcal{L}}_{\mathrm{Hopf}_{GG}}$. This is a known feature of such terms
which have been extensively studied to explain $CP$-violation, both in Abelian and Non-Abelian gauge theories (see for instance~\cite{Witten,PS} and references therein).

We have listed all the possible lower order candidate terms to build our action.
We also need to study how these several candidate terms behave under electromagnetic duality:
\be
\ba{rcl}
{\mathcal{L}}_{\mathrm{Maxwell}_{FF}}&\to &\cos^2(\theta){\mathcal{L}}_{\mathrm{Maxwell}_{FF}}+\sin^2(\theta){\mathcal{L}}_{\mathrm{Maxwell}_{GG}}+2\cos(\theta)\sin(\theta){\mathcal{L}}_{\mathrm{Maxwell}_{FG}}\\[5mm]
{\mathcal{L}}_{\mathrm{Maxwell}_{GG}}&\to &\sin^2(\theta){\mathcal{L}}_{\mathrm{Maxwell}_{FF}}+\cos^2(\theta){\mathcal{L}}_{\mathrm{Maxwell}_{GG}}-2\cos(\theta)\sin(\theta){\mathcal{L}}_{\mathrm{Maxwell}_{FG}}\\[5mm]
{\mathcal{L}}_{\mathrm{Hopf}_{FG}}&\to &\sin\theta\cos\theta\left({\mathcal{L}}_{\mathrm{Hopf}_{GG}}-{\mathcal{L}}_{\mathrm{Hopf}_{FF}}\right)+(\cos^2\theta-\sin^2\theta){\mathcal{L}}_{\mathrm{Hopf}_{FG}}\\[5mm]
{\mathcal{L}}_{\mathrm{Maxwell}_{FG}}&\to &\sin\theta\cos\theta\left({\mathcal{L}}_{\mathrm{Maxwell}_{GG}}-{\mathcal{L}}_{\mathrm{Maxwell}_{FF}}\right)+(\cos^2\theta-\sin^2\theta){\mathcal{L}}_{\mathrm{Maxwell}_{FG}}\\[5mm]
{\mathcal{L}}_{\mathrm{Hopf}_{FF}}&\to &\cos^2(\theta){\mathcal{L}}_{\mathrm{Hopf}_{FF}}+\sin^2(\theta){\mathcal{L}}_{\mathrm{Hopf}_{GG}}+2\cos(\theta)\sin(\theta){\mathcal{L}}_{\mathrm{Hopf}_{FG}}\\[5mm]
{\mathcal{L}}_{\mathrm{Hopf}_{GG}}&\to &\sin^2(\theta){\mathcal{L}}_{\mathrm{Hopf}_{FF}}+\cos^2(\theta){\mathcal{L}}_{\mathrm{Hopf}_{GG}}-2\cos(\theta)\sin(\theta){\mathcal{L}}_{\mathrm{Hopf}_{FG}}
\ea
\lb{dual_Hopf}
\ee

We are now ready to build an action that describes electromagnetism with two gauge fields. Demanding
the action to be $P$ and $T$ invariant we are left only with the terms listed in~(\ref{L_inv}).
So we conclude that the most standard action that explicitly depends on two gauge fields
must be a combination of ${\mathcal{L}}_{\mathrm{Maxwell}_{FF}}$ and ${\mathcal{L}}_{\mathrm{Maxwell}_{GG}}$.
We will call this action the minimal action~\cite{Sing_03,action_01,action_02}
\bea
S_{\mathrm{Min}_+}&=&-\int_M\sqrt{-g}\left[\frac{1}{4\,e^2}F_{\mu\nu}F^{\mu\nu}+\frac{1}{4\,g^2}G_{\mu\nu}G^{\mu\nu}\right]\lb{S_min_+}\\
S_{\mathrm{Min}_-}&=&-\int_M\sqrt{-g}\left[\frac{1}{4\,e^2}F_{\mu\nu}F^{\mu\nu}-\frac{1}{4\,g^2}G_{\mu\nu}G^{\mu\nu}\right]\ .\lb{S_min_-}
\eea
We note that from the electric and magnetic fields definition~(\ref{EB}) both Maxwell terms must have the same numerical
factor (up to the relative sign). The standard would be to consider both with the same sign in order to
have the same quantum structure in both sectors, however for completeness we consider both cases.
These actions imply the existence of two electric and two magnetic fields as we will discuss in detail
in section~\ref{sec.fields}. Instead we expect to have only one electric and one magnetic field
such that the group charge flux of one $U(1)$ is of the same nature of the flux of topological
charge of the other $U(1)$ as implied by the field definitions~(\ref{EB}). As a \textit{weaker}
but valid argument we note that the pure gauge sectors are completely decoupled, \textit{a priori} one would
expect that some sort of mixing (meaning coupling) between the two sectors exist that, at least, accomplishes
the coupling of topological flux with group charge fluxes. We consider these argument as a drawback of
the minimal actions.

From the above arguments we are further considering the remaining allowed term that preserves $T$ and $P$,
the cross Hopf term. We call these actions the maximal actions 
\bea
\displaystyle S^{\hat{\epsilon}}_{\mathrm{Max}_+}&=&\displaystyle-\int_M\left[\frac{\sqrt{-g}}{4\,e^2}F_{\mu\nu}F^{\mu\nu}+\frac{\sqrt{-g}}{4\,g^2}G_{\mu\nu}G^{\mu\nu}-\frac{\hat{\epsilon}}{4\,e\,g}\epsilon^{\mu\nu\rho\delta}F_{\mu\nu}G_{\rho\delta}\right]\lb{S_max_+}\\[5mm]
\displaystyle S^{\hat{\epsilon}}_{\mathrm{Max}_-}&=&\displaystyle-\int_M\left[\frac{\sqrt{-g}}{4\,e^2}F_{\mu\nu}F^{\mu\nu}-\frac{\sqrt{-g}}{4\,g^2}G_{\mu\nu}G^{\mu\nu}-\frac{\hat{\epsilon}}{4\,e\,g}\epsilon^{\mu\nu\rho\delta}F_{\mu\nu}G_{\rho\delta}\right]\lb{S_max_-}
\eea
The cross Hopf term couples the flux of the group charges ($F^{0i}$ and $G^{0i}$) with the flux
of the topological charge ($G_{ij}$ and $F_{ij}$ respectively) of the two different $U(1)$'s. As
we are going to show in the next section~\ref{sec.fields} the only action that can be defined
using only one electric and one magnetic field is $S^{\hat{\epsilon}}_{\mathrm{Max}_-}$ as given
by~(\ref{S_max_-}). $\hat{\epsilon}=\pm 1$ sets the relative sign of the Hopf term and will be
relevant in the definition of the physical fields as we will show in detail.

We also note that when both Maxwell terms have the same sign the minimal action $S_{\mathrm{Min}_+}$
as given by~(\ref{S_min_+}) is invariant under electromagnetic duality, so for an action of this form
we have elevated the duality to a symmetry~\cite{new_01,Sing_04}. Also the respective Hamiltonian will be
duality invariant. However this is not necessarily a good feature. As we already pointed out,
standard duality does not preserve either $P$ or $T$ and this fact is not explicit on the action
$S_{\mathrm{Min}_-}$, neither can be on a duality invariant action. This argument is not completly
close because one can consider the duality angle parameter $\theta$ to be a pseudo-scalar~\cite{Sing_03}.
However another physical argument is that due to Dirac quantization condition~\cite{Dirac} ($eg=n$) we
have that the $A$ field obeys a weak coupling regime while the $C$ field obeys a strong coupling regime.
Then we expect that the energy (Hamiltonian) not to be conserved under a duality rotation. 
For these reasons a duality symmetric action does not look like a good choice.
The remaining actions, the maximal actions~(\ref{S_max_-}) and~(\ref{S_max_+}) and the minimal action~(\ref{S_min_-}),
are not invariant under duality but they are $P$ and $T$ invariant. Furthermore a duality
transformation does not preserve $P$ and $T$ invariance as can explicitly be seen from~(\ref{dual_Hopf}).
This is actually a good feature, duality explicitly breaks $P$ and $T$ at the level of the action as 
expected from~(\ref{dualEB_P}) and~(\ref{dualEB_T}) and the respective Hamiltonians are not invariant under duality.

Finally the as we show in detail in the next subsection, only $S^{\hat{\epsilon}}_{\mathrm{Max}_-}$
is compatible with the existence of only one electric and one magnetic physical fields.

\subsection{Physical Electric and Magnetic Fields\lb{sec.fields}}

Due to have introduced a second gauge field $C$ we have now twice the degrees of freedom than usual electromagnetism.
Accordingly we expect to have as well a new interaction such that generally we have two \textit{electric fields} and two
\textit{magnetic fields}. From a theoretical point of view this is standard, each of the gauge fields
carry a different kind of interaction. Nevertheless we are led to the question if both definitions are
physical fields or not. Here we will show that for both the minimal actions~(\ref{S_min_+}) and~(\ref{S_min_-})
and the maximal action~(\ref{S_max_+}) we have indeed four physical fields (two electric and two magnetic),
while for the maximal action~(\ref{S_max_-}) we have only two physical fields (one electric and one magnetic).

Let us consider the generic definitions of electric and magnetic fields
corresponding to the gauge fields $A$ and $C$
\be
\ba{rclcrcl}
E_A^i&=&\displaystyle\frac{1}{e}\,F^{0i}&\ \ \ \ \ \ &E_C^i&=&\displaystyle-\frac{1}{2g}\,\epsilon^{ijk}G_{jk}\\[5mm]
B_A^i&=&\displaystyle\frac{1}{2e}\,\epsilon^{ijk}F_{jk}& &B_C^i&=&\displaystyle\frac{1}{g}\,G^{0i}\ .
\ea
\lb{EB_AC}
\ee
We note that the definitions of electric and magnetic fields for $C$ are reversed to the
ones of the $A$ field and for reasons that will become clear in the remaining of this section
we consider a minus sign in the definition of $\vb{E_C}$. Both in order to define the electric and magnetic
fields accordingly to~(\ref{EB}) and to preserve the properties of the fields in relation to
the discrete symmetries, i.e. the electric field is a vector and the magnetic field is a pseudo-vectors.
Then we take the following linear combinations of the above definitions~(\ref{EB_AC})
\be
\ba{rclcl}
E_+^i&=&\displaystyle\frac{1}{2}\left(E_A^i+E_C^i\right)&=&\displaystyle+\frac{1}{2e}\,F^{0i}-\frac{1}{4g}\,\epsilon^{ijk}G_{jk}\\[5mm]
B_+^i&=&\displaystyle\frac{1}{2}\left(B_A^i+B_C^i\right)&=&\displaystyle+\frac{1}{2g}\,G^{0i}+\frac{1}{4e}\,\epsilon^{ijk}F_{jk}\\[5mm]
E_-^i&=&\displaystyle\frac{1}{2}\left(E_A^i-E_C^i\right)&=&\displaystyle+\frac{1}{2e}\,F^{0i}+\frac{1}{4g}\,\epsilon^{ijk}G_{jk}\\[5mm]
B_-^i&=&\displaystyle\frac{1}{2}\left(B_A^i-B_C^i\right)&=&\displaystyle-\frac{1}{2g}\,G^{0i}+\frac{1}{4e}\,\epsilon^{ijk}F_{jk}\ .
\ea
\lb{EB_pm}
\ee
To consider these combination is the approach of several authors that
consider only $\vb{E_+}$ and $\vb{B_+}$ as physical fields~\cite{Sing_02,action_01}.
As already explained in the introduction the main motivation is to
achieve a generalized description of electromagnetism with both electric and magnetic particles~\cite{CF}.
Also these combinations are used to implement an explicit electromagnetic duality
between the two sectors in terms of two distinct gauge fields as we explained in section~\ref{sec.dual}.

For what follows we will need the identities
\be
\ba{rcccl}
\displaystyle\frac{1}{2\,e^2}F^{0i}F_{0i}&=&\displaystyle\frac{1}{2}\left(E_+^i+E_-^i\right)\left(E_+^i+E_-^i\right)&=&\displaystyle\frac{1}{2}\left(+E_+^iE_+^i+E_-^iE_-^i+2\,E_+^iE_-^i\right)\\[5mm]
\displaystyle\frac{1}{4\,e^2}F^{ij}F_{ij}&=&\displaystyle-\frac{1}{2}\left(B_+^i+B_-^i\right)\left(B_+^i+B_-^i\right)&=&\displaystyle\frac{1}{2}\left(-B_+^iB_+^i-B_-^iB_-^i-2\,B_+^iB_-^i\right)\\[5mm]
\displaystyle\frac{1}{2\,g^2}G^{0i}G_{0i}&=&\displaystyle\frac{1}{2}\left(B_+^i-B_-^i\right)\left(B_+^i-B_-^i\right)&=&\displaystyle\frac{1}{2}\left(+B_+^iB_+^i+B_-^iB_-^i-2\,B_+^iB_-^i\right)\\[5mm]
\displaystyle\frac{1}{4\,g^2}G^{ij}G_{ij}&=&\displaystyle-\frac{1}{2}\left(-E_+^i+E_-^i\right)\left(-E_+^i+E_-^i\right)&=&\displaystyle\frac{1}{2}\left(-E_+^iE_+^i-E_-^iE_-^i+2\,E_+^iE_-^i\right)\\[5mm]
\displaystyle\frac{1}{2\,e\,g}\epsilon^{0ijk}F_{0i}G_{jk}&=&\displaystyle \left(E_+^i+E_-^i\right)\left(-E_+^i+E_-^i\right)&=&\displaystyle -E_+^iE_+^i+E_-^iE_-^i\\[5mm]
\displaystyle\frac{1}{2\,e\,g}\epsilon^{0ijk}G_{0i}F_{jk}&=&\displaystyle \left(B_+^i+B_-^i\right)\left(B_+^i-B_-^i\right)&=&\displaystyle +B_+^iB_+^i-B_-^iB_-^i
\ea
\lb{defs}
\ee
where the minus sign and the factor of $2$ in the second and forth lines are due to the contraction of the indices of the antisymmetric
tensor, i.e. $\epsilon^{0jki}\epsilon^{\ \ \ \ i'}_{0jk}=-2\delta^{ii'}$.

Let us consider both the minimal actions~(\ref{S_min_+}) and~(\ref{S_min_-}) and the maximal actions~(\ref{S_max_+})
and~(\ref{S_max_-}) and rewrite the respective Lagrangians in terms of the above combinations~(\ref{EB_pm}) using the
identities~(\ref{defs}). We obtain that
\be
\ba{rcl}
\displaystyle{\mathcal{L}}_{\mathrm{Min}_+}&=&\displaystyle -2\left(E_+^iE_-^i-B_+^iB_-^i\right)\\[5mm]
\displaystyle{\mathcal{L}}_{\mathrm{Min}_-}&=&\displaystyle-\left(E_+^iE_+^i+E_-^iE_-^i-B_+^iB_+^i-B_-^iB_-^i\right)\\[5mm]
\displaystyle{\mathcal{L}}^{\hat{\epsilon}}_{\mathrm{Max}_+}&=&\displaystyle -2\left(E_+^iE_-^i-B_+^iB_-^i\right)-\hat{\epsilon}\left(E_+^iE_+^i-E_-^iE_-^i-B_+^iB_+^i+B_-^iB_-^i\right)\\[5mm]
\displaystyle{\mathcal{L}}^{\hat{\epsilon}}_{\mathrm{Max}_-}&=&\displaystyle -2\left(E_{\hat{\epsilon}}^iE_{\hat{\epsilon}}^i-B_{\hat{\epsilon}}^iB_{\hat{\epsilon}}^i\right)\ .
\ea
\lb{L_EB}
\ee
Here we use the usual convention in classical
electrodynamics~\cite{Jackson}, we sum over repeated indices $i$ that are considered always upstairs
such that the metric is no longer explicit, because $E^i=-E_i$ we have that $\vb{E}^2=-E^iE_i=E^iE^i\geq 0$.
The indices of the electric and magnetic fields correspond to $\pm$ depending respectively on what the
choice $\hat{\epsilon}=\pm 1$ is taken.

We readily conclude that the only action that can be written in terms of only 
two fields ($\vb{E_+}$ and $\vb{B_+}$ for $\hat{\epsilon}=+1$ or $\vb{E_-}$ and $\vb{B_-}$
for $\hat{\epsilon}=-1$) is the maximal action~(\ref{S_max_-}). This is only possible if the
cross Hopf term is present and the two Maxwell terms have opposite sign. We also note that the
choice of $\hat{\epsilon}$ is relevant to the physical field definitions.

As a worm up for what follows we can argue that, after our field redefinition~(\ref{EB_pm}),
the form of the maximal Lagrangian ${\mathcal{L}}^{\hat{\epsilon}}_{\mathrm{Max}_-}$
as given in the last line of~(\ref{L_EB}) is essentially the same as the one of standard electromagnetism,
therefore we have a very strong indication that, for this Lagrangian, the physical fields are
\be
\ba{rcl}
E^i&=&\displaystyle \frac{1}{e}F^{0i}-\frac{\hat{\epsilon}}{2g}\epsilon^{0ijk}G_{jk}\\[5mm]
B^i&=&\displaystyle\frac{\hat{\epsilon}}{g}G^{0i}+\frac{1}{2e}\epsilon^{0ijk}F_{jk}\ .
\ea
\lb{EB_phys}
\ee
This is only possible for the maximal Lagrangian ${\mathcal{L}}^{\hat{\epsilon}}_{\mathrm{Max}_-}$.

To show that this is indeed the case we will formalize this argument and analyse the four possible actions.
Let us compute the equations of motion for the actions and check which fields appear in them. We will
properly discuss how to couple each type of current densities to both the gauge fields in the next section, for the
moment being let us assume the standard minimal coupling
\be
S_{\mathrm{Sources},\mathrm{Min}_\pm}=-\frac{1}{e}\int A_\mu J_e^\mu\pm\frac{1}{g}\int C_\mu J_g^\mu
\lb{sources_min}
\ee
where the $\pm$ correspond to the relative sign between the Maxwell terms.

For the minimal actions we have that the equations of motion are
\be
\left\{
\ba{rcl}
\displaystyle\frac{1}{e}\partial_\mu F^{\mu\nu}&=&J^\mu_e\\[5mm]
\displaystyle\frac{1}{g}\partial_\mu G^{\mu\nu}&=&J^\mu_g
\ea
\right.
\ \Leftrightarrow\ 
\left\{
\ba{rcl}
\nabla.\vb{E}_A&=&\rho_e\\[5mm]
\nabla.\vb{B}_C&=&\rho_g\\[5mm]
\dot{\vb{B}}_C+\nabla\times \vb{E}_C&=&-\vb{J}_g\\[5mm]
\dot{\vb{E}}_A-\nabla\times \vb{B}_A&=&-\vb{J}_e\ .
\ea
\right.
\ee
The electric and magnetic equations are completely decoupled and we have two electric and two magnetic fields.
Also in addition to this equation we have the Bianchi identities for each gauge field.
There is a way to couple both sector by using non homogeneous Bianchi identities, for that consider
non regular gauge fields such that we have the respective Bianchi identities $dF=*\tilde{J}_g$ and
$dG=*\tilde{J}_e$. Then by an appropriate combination of the equations of motion with the Bianchi
identities we obtain $d(*F-G)=*J_e-*\tilde{J}_e$ and $d(F+*G)=*J_g+*\tilde{J}_g$
which correspond to the generalized Maxwell equations~(\ref{Maxwell}) with the current densities
changed from $J_e\to J_e-\tilde{J}_e$ and $J_g\to J_g+\tilde{J}_g$. Here $*$ denotes the usual Hodge
duality operation and we used form notation for compactness. There are two drawbacks for this approach,
first the current densities are no longer the ones which minimally couple to the gauge fields at the level of the action
and secondly, the identification of the topological charge fluxes with the group charge fluxes of different gauge
groups is imposed (by hand) not emerging naturally from the action. These problems are solved by using the
maximal action with opposite signs for the Maxwell terms as given by~(\ref{S_max_-}).

In order to analise the maximal action~(\ref{S_max_-}) we note that the above procedure of redefinition
of fields~(\ref{EB_pm}) and rewriting the Lagrangians in terms of the redefined fields~(\ref{L_EB}) is
equivalent to rewriting the Lagrangian in terms of the new 2-form \textit{gauge connections}~\cite{Sing_03}
\be
\ba{rclcl}
\left({\mathcal{F}}_+^{\hat{\epsilon}}\right)^{\mu\nu}&=&\displaystyle\frac{1}{2}\left(F+\hat{\epsilon}*G\right)^{(\mu\nu)}&=&\displaystyle\frac{1}{2e}F^{\mu\nu}+\frac{\hat{\epsilon}}{4g}\epsilon^{\mu\nu\rho\delta}G_{\rho\delta}\\[5mm]
\left({\mathcal{F}}_-^{\hat{\epsilon}}\right)^{\mu\nu}&=&\displaystyle\frac{1}{2}\left(F-\hat{\epsilon}*G\right)^{(\mu\nu)}&=&\displaystyle\frac{1}{2e}F^{\mu\nu}-\frac{\hat{\epsilon}}{4g}\epsilon^{\mu\nu\rho\delta}G_{\rho\delta}.
\ea
\lb{F_pm}
\ee
or their Hodge duals
\be
\ba{rclcl}
\left({\mathcal{G}}_+^{\hat{\epsilon}}\right)^{\mu\nu}&=&-*\left({\mathcal{F}}_-^{\hat{\epsilon}}\right)^{(\mu\nu)}&=&\displaystyle\frac{\hat{\epsilon}}{2g}G^{\mu\nu}+\frac{1}{4g}\epsilon^{\mu\nu\rho\delta}F_{\rho\delta}\\[5mm]
\left({\mathcal{G}}_-^{\hat{\epsilon}}\right)^{\mu\nu}&=&-*\left({\mathcal{F}}_+^{\hat{\epsilon}}\right)^{(\mu\nu)}&=&\displaystyle\frac{\hat{\epsilon}}{2e}G^{\mu\nu}-\frac{1}{4g}\epsilon^{\mu\nu\rho\delta}F_{\rho\delta}.
\ea
\lb{G_pm}
\ee

Then the maximal Lagrangian~(\ref{S_max_-}) is rewritten in both equivalent expressions as
\be
{\mathcal{L}}^{\hat{\epsilon}}_{\mathrm{Max}_-}=-\left({\mathcal{F}}_+^{\hat{\epsilon}}\right)^{\mu\nu}\left({\mathcal{F}}_+^{\hat{\epsilon}}\right)_{\mu\nu}=+\left({\mathcal{G}}_-^{\hat{\epsilon}}\right)^{\mu\nu}\left({\mathcal{G}}_-^{\hat{\epsilon}}\right)_{\mu\nu}\ .
\lb{L_F_pm}
\ee
where we used the Hodge duality property $**G=-G$ for 2-forms $G$ in Lorentzian $4D$ manifolds.
This is basically the reason why in~(\ref{EB_AC}) we defined $E_C^i=-\epsilon^{ijk}G^{jk}$ with
a minus sign~\cite{Sing_02,action_01}. We note that these two ways of rewriting are algebraically equivalent.
However physically they have an important meaning, we can have both a electric and a magnetic description
of the theory. This is seen in the equations of motion. Upon variation of the maximal action with respect
to $A$ and $C$ we obtain
\be
\ba{rcl}
\partial_\mu\left({\mathcal{F}}_+^{\hat{\epsilon}}\right)^{\mu\nu}&=&J^\nu_e\\[5mm]
\partial_\mu\left({\mathcal{G}}_-^{\hat{\epsilon}}\right)^{\mu\nu}&=&\hat{\epsilon}J^\nu_g
\ea
\lb{EOM_FG}
\ee
which indeed correspond to the generalize Maxwell equations~(\ref{Maxwell}) and are
expressed only in terms of the fields $\vb{E}$ and $\vb{B}$ as given by~(\ref{EB_phys}).
So these must be the physical fields! This is only possible for the maximal action.
As for ${\mathcal{L}}^{\hat{\epsilon}}_{\mathrm{Max}_+}$ corresponding to the action~(\ref{S_max_+})
this construction is not possible, we obtain that
\be
{\mathcal{L}}^{\hat{\epsilon}}_{\mathrm{Max}_+}=-\frac{1}{2}\left({\mathcal{F}}_+^{\hat{\epsilon}}\right)^{\mu\nu}\left({\mathcal{F}}_+^{\hat{\epsilon}}\right)_{\mu\nu}+\frac{1}{2}\left({\mathcal{F}}_-^{\hat{\epsilon}}\right)^{\mu\nu}\left({\mathcal{F}}_-^{\hat{\epsilon}}\right)_{\mu\nu}-\left({\mathcal{F}}_+^{\hat{\epsilon}}\right)^{\mu\nu}\left({\mathcal{F}}_-^{\hat{\epsilon}}\right)_{\mu\nu}
\ee
such that we need two distinct \textit{gauge connections} in order to define it, hence as expected four
physical fields.

One must be careful with the way we couple the source to both gauge fields
depending on the choice of $\hat{\epsilon}$ due to the definitions of the physical fields and
the current sign in the second line of~(\ref{EOM_FG}). We will discuss this issue in detail in
the next section~\ref{sec.currents}.

There is a subtlety here. The reader may by now be recalling the Bianchi identities
(or homogeneity conditions for Abelian gauge fields)
on the gauge connection and claiming that as usual for topological terms
the variation
\be
\delta{\mathcal{L}}_{\mathrm{Hopf}_{FG}}=\frac{1}{2\,e\,g}\epsilon^{\mu\nu\delta\rho}(\partial_\nu G_{\delta\rho}\delta A_\mu+\partial_\nu F_{\delta\rho}\delta C_\mu)
\ee
should be always null and does not contribute to the equations of motion. This is true for regular fields,
however as already mentioned in the first section and in the analysis of the minimal actions,
if non-regular gauge fields are allowed then this contribution to the equations of motion is not null
everywhere and must be taken in account and~(\ref{EOM_FG}) are actually the correct ones.
By discontinuities we mean that $\partial_\mu\partial_\nu C\neq\partial_\nu\partial_\mu C$.
Allowing for corrections to the Bianchi identities allows for the inclusion of magnetic
charge in standard electromagnetism (with only one $U(1)$ gauge field) and is in the basis of
the original construction that originates the Dirac string~\cite{Dirac} or the equivalent
non-trivial fiber-bundle of Wu and Yang~\cite{YW}. We present this argument only to show that
algebraically~(\ref{EOM_FG}) are correct, we don't need to \textit{necessarily}
have these discontinuities to describe both electric and magnetic charge as long as we work
with two distinct gauge fields. However we show in~\cite{outro} that in order
to have effective theories obtained from the maximal action only with one gauge field
we still have discontinuities, but the discontinuities will be present on the
extra field (instead of the physical field of the effective theory as in\cite{Dirac,YW}).)

An important result here is that for the maximal action the topological fluxes
of one $U(1)$ are identified with the charge fluxes of the other $U(1)$ as desired
for the existence of only one electric and one magnetic physical fields.
We must stress that this does not imply that we are constraining the fundamental fields $A$ and $C$, we
are maintaining the same degrees of freedom. We have 4 physical degrees of freedom (2 for each of the gauge fields
$A$ and $C$) which are still maintained in the electric and magnetic fields (again 2 for each of the
fields $E$ and $B$). In standard electromagnetism with only one gauge field there is
only 2 degrees of freedom. The interpretation in terms of the fields is quite interesting.
For each of the $U(1)$ fields the 2 physical degrees of freedom correspond to the transverse modes
while the longitudinal modes are not physical and do not constitute physical degrees of freedom.
When combining the gauge connections as in~(\ref{F_pm}) the degrees of freedom of the second gauge
field $C$ are combined with the degrees of freedom of the original gauge field $A$ in such a way
that they play the role of two Longitudinal modes of the gauge field $A$, simply we have
now two longitudinal modes instead of a single one as is usual in theories with massive photons. These
degrees of freedom constitute here physical degrees of freedom and are due to the inclusion
of a second $U(1)$ gauge group.

Our discussion would not be complete without discussing the canonical variables.
We do so next and also discuss briefly the expression for the Hamiltonians corresponding
to the minimal and maximal actions.

\subsection{Canonical Variables and Hamiltonian Formulation}

The canonical momenta for the minimal actions~(\ref{S_min_+}) and~(\ref{S_min_-}) are
\be
\ba{rclcl}
\pi_{A,\mathrm{Min}}^i&=&\displaystyle\frac{1}{e^2}F^{0i}&=&\displaystyle\frac{1}{e}E_A^i\\[5mm]
\pi_{C,\mathrm{Min}_\pm}^i&=&\displaystyle\pm \frac{1}{g^2}G^{0i}&=&\displaystyle\pm\frac{1}{g}B_C^i\\[5mm]
\ea
\ee
where the $\pm$ refers respectively to ${\mathcal{L}}_{\mathrm{Min}_+}$ (the $+$ sign) and
${\mathcal{L}}_{\mathrm{Min}_-}$ (the $-$ sign). This means that the canonical momenta are
each of the $U(1)$ group charge fluxes. The Hamiltonian depends on both gauge sectors
but each of them are completely decoupled
\be
\ba{rcl}
{\mathcal{H}}_{{\mathrm{Min}}_\pm}&=&\displaystyle\frac{1}{2}\left(e^2\,\pi_{A,\mathrm{Min}}^i\pi_{A,\mathrm{Min}}^i+\frac{1}{2\,e^2}F_{ij}F_{ij}\right)\\[5mm]
                 & &\displaystyle\pm \frac{1}{2}\left(g^2\,\pi_{C,\mathrm{Min}_\pm}^i\pi_{C,\mathrm{Min}_\pm}^i+\frac{1}{2\,g^2}G_{ij}G_{ij}\right)
\ea
\ee
such that the Hilbert space factorizes into states carrying charge fluxes of both gauge sectors.
The topological charge fluxes are present only trough the potential $F_{ij}F_{ij}$ and $G_{ij}G_{ij}$
as in standard electromagnetism. So basically we have two distinct copies of standard electromagnetism
and no interaction terms between the two sectors.

The canonical momenta for the maximal action~(\ref{S_max_+}) and~(\ref{S_max_-})  are
\be
\ba{rclcl}
\left(\pi_{A,\mathrm{Max}_\pm}^{\hat{\epsilon}}\right)^i&=&\displaystyle\frac{1}{e^2}F^{0i}\pm\frac{\hat{\epsilon}}{2\,e\,g}\epsilon^{ijk}G_{jk}&=&\displaystyle+\frac{2}{e}E_{(\pm\hat{\epsilon})}^i\\[5mm]
\left(\pi_{C,\mathrm{Max}_\pm}^{\hat{\epsilon}}\right)^i&=&\displaystyle\pm\frac{1}{g^2}G^{0i}-\frac{\hat{\epsilon}}{2\,e\,g}\epsilon^{ijk}F_{jk}&=&\displaystyle-\frac{2\,\hat{\epsilon}}{g}B_\pm^i\\[5mm]
\ea
\ee
where the $\pm$ refers respectively to ${\mathcal{L}}_{\mathrm{Max}_+}$ (the $+$ sign) and
${\mathcal{L}}_{\mathrm{Max}_-}$ (the $-$ sign). In the electric field the subscript $(\pm\hat{\epsilon})$
means the product of $\pm 1$ by $\hat{\epsilon}$. The canonical momenta coincide up to
constants with the physical electric and magnetic fields, this is a good indication that
indeed, also at quantum level, we can have the correct identifications between group charge and
topological charge fluxes from the opposite $U(1)$'s.

After a straight forward computation we obtain the following Hamiltonians
\be
\ba{rl}
{\mathcal{H}}^{\hat{\epsilon}}_{\mathrm{Max}_+}=&\displaystyle+\frac{e^2}{2}\left(\pi_{A,\mathrm{Max}_+}^i+\frac{\hat{\epsilon}}{2\,e\,g}\epsilon^{ijk}G_{jk}\right)\left(\pi_{A,\mathrm{Max}_+}^i-\frac{\hat{\epsilon}}{2\,e\,g}\epsilon^{ijk}G_{jk}\right)\\[5mm]
              &\displaystyle+\frac{g^2}{2}\left(\pi_{C,\mathrm{Max}_+}^i+\frac{\hat{\epsilon}}{2\,e\,g}\epsilon^{ijk}F_{jk}\right)\left(\pi_{C,\mathrm{Max}_+}^i-\frac{\hat{\epsilon}}{2\,e\,g}\epsilon^{ijk}F_{jk}\right)\\[5mm]
              &\displaystyle+\frac{\hat{\epsilon}}{2\,e\,g}\epsilon^{ijk}\left(\pi_{A,\mathrm{Max}_+}^iG_{jk}+\pi_{C,\mathrm{Max}_+}^iF_{jk}\right)-\frac{3}{4\,e^2}F_{ij}F_{ij}+\frac{5}{4g^2}G_{ij}G_{ij}
\ea
\lb{H_max_+}
\ee
and
\be
\ba{rl}
{\mathcal{H}}^{\hat{\epsilon}}_{\mathrm{Max}_-}=&\displaystyle+\frac{e^2}{2}\left(\pi_{A,\mathrm{Max}_-}^i+\frac{\hat{\epsilon}}{2\,e\,g}\epsilon^{ijk}G_{jk}\right)\left(\pi_{A,\mathrm{Max}_-}^i-\frac{\hat{\epsilon}}{2\,e\,g}\epsilon^{ijk}G_{jk}\right)\\[5mm]
              &\displaystyle-\frac{g^2}{2}\left(\pi_{C,\mathrm{Max}_-}^i+\frac{\hat{\epsilon}}{2\,e\,g}\epsilon^{ijk}F_{jk}\right)\left(\pi_{C,\mathrm{Max}_-}^i-\frac{\hat{\epsilon}}{2\,e\,g}\epsilon^{ijk}F_{jk}\right)\\[5mm]
              &\displaystyle+\frac{\hat{\epsilon}}{2\,e\,g}\epsilon^{ijk}\left(\pi_{A,\mathrm{Max}_-}^iG_{jk}-\pi_{C,\mathrm{Max}_-}^iF_{jk}\right)+\frac{5}{4\,e^2}F_{ij}F_{ij}-\frac{5}{4g^2}G_{ij}G_{ij}\ .
\ea
\lb{H_max_-}
\ee
The first lines of both these equations are interpreted as usual with $a_+^ia_-^i$ where $a_\pm$ are creation
and annihilation operators of electric excitations and the second lines correspond to $b_+^ib_-^i$ where $b_\pm$
are creation and annihilation operators of magnetic excitations. The third lines contain a generalized
\textit{angular momenta} term between the two gauge sectors and the potentials $F_{ij}F_{ij}$ and $G_{ij}G_{ij}$.
We note that the potential terms have non-standard factors and opposite signs in both Hamiltonians. In particular the
factors for the potentials in ${\mathcal{H}}^{\hat{\epsilon}}_{\mathrm{Max}_+}$ as given in~(\ref{H_max_+}) have
different weights (i.e. besides the opposite sign have different numerical factors independently of the coupling
constants) while they have the same numerical weight for ${\mathcal{H}}^{\hat{\epsilon}}_{\mathrm{Max}_-}$ as
given in~(\ref{H_max_-}). This is also a good indication that the Maximal action~(\ref{S_max_-}) is the correct
one since there is no reason for the potentials $F_{ij}F_{ij}$ and $G_{ij}G_{ij}$ having different numerical weights
(besides the coupling constants).

The Hilbert space is not generally factorizable, the states should only be factorizable for states that have
null eigenvalues of the generalized angular momenta.

The main problem in quantizing this theory is that the $b_+^ib_-^i$ has the opposite sign (than the standard fields)
and, using the usual commutation relations for the $C$ field, makes the existence of negative energy states possible.
In order to solve this issue the standard way out is to consider anti-commutation
relations for the $C$ gauge sector~\cite{PS}. In this case we are in presence of a ghost field~\cite{Linde},
not a standard boson. An alternative approach is to consider some mechanism that allows to quantize only
the electric sector as done in~\cite{quant}. As such examples we have Phantom matter in
cosmology~\cite{Phantom} where such fields are considered at classical level (i.e. we may consider them
to be a collective field, meaning a statistical effective field) in inflationary models. Also we can consider
a dynamical symmetry breaking mechanism~\cite{outro}, a possible application is considered in~\cite{Tito_01,Tito_02}
as a way to generate a Proca mass for the usual photon.

We are not discussing any further the quantization procedure here.

\section{Inclusion of Current Densities\lb{sec.currents}}

In here we are analysing in detail the current densities coupling to both the gauge fields 
for the Maximal action~(\ref{S_max_-}) and accordingly derive the Lorentz force with
both gauge fields.

\subsection{Current Coupling Terms}

Concerning the inclusion of currents let us consider the standard action
\be
S^{\hat{\epsilon}}_{\mathrm{Sources},\mathrm{Max}_-}=-\int_M\left[\frac{1}{e}A_\mu J_e^\mu-\frac{\hat{\epsilon}}{g}C_\mu J_g^\mu\right]
\lb{S_Sources}
\ee
where we have the $\hat{\epsilon}$ correctly sets the current sign in the generalized Maxwell equations~(\ref{EOM_FG}).
This action is both $P$ and $T$ invariant but it is not invariant under electromagnetic
duality rotation. Under a duality rotation we effectively couple each current density with both
gauge fields obtaining violating terms.

From the discussion of the last section we concluded that the physical electric and magnetic
fields are given by~(\ref{EB_phys}). So each of the currents need to couple in some way to both $U(1)$
gauge fields. The question is how to do it maintaining $P$ and $T$ symmetries and having the
variation of the action with respect to space-time coordinates holding the
Lorentz force defined in terms of the fields~(\ref{EB}). We note that~(\ref{S_Sources}) is not
enough since it holds that we would have two Lorentz forces, one for each $U(1)$'s
in terms of the decoupled fields as given in~(\ref{EB_AC}).

The way out is to consider the dual fields $\tilde{A}$ and $\tilde{C}$ defined in terms of the original gauge fields
by the differential equations
\be
\left\{
\ba{rcl}
\tilde{F}&=&*F\\[5mm]
\tilde{G}&=&*G
\ea\right.\ \ \Leftrightarrow\ \ 
\left\{
\ba{rcl}
d\tilde{A}&=&*dA\\[5mm]
d\tilde{C}&=&*dC
\ea\right.
\lb{dual_fields}
\ee
where again $*$ denotes the Hodge duality operation. We note that the dual fields have only longitudinal modes,
so by dual we mean that we are exchanging transverse modes in $A$ and $C$ by longitudinal modes in $\tilde{A}$
and $\tilde{C}$. So the extra action for the current densities read
\be
S_{\mathrm{Dual\ Sources},\mathrm{Max}_-}=+\int_M\left[\frac{\hat{\epsilon}}{g}\tilde{C}_\mu J_e^\mu+\frac{1}{e}\tilde{A}_\mu J_g^\mu\right]
\lb{S_Sources_dual}
\ee
Both terms are $P$ and $T$ invariant because $\tilde{A}$ and $\tilde{C}$ are, respectively, a
pseudo-vector and a vector due to~(\ref{dual_fields}). Again the sign choice is not arbitrary, we already fixed it
in order to obtain the correct Lorentz forces. Electromagnetic duality couples both current densities with
both gauge fields $\tilde{A}$ and $\tilde{C}$ such that it induces $P$ and $T$ violating terms.
For this action we indeed have that the group charges of each $U(1)$ (given by the $J$'s)
are coupled to the topological charges of the other $U(1)$ (in terms of $\tilde{A}$ and $\tilde{C}$).
This is what is expressed in the definition of the dual fields as given by~(\ref{dual_fields}).
Also there are a couple of very important points we must address. These terms do not contribute
to the equations of motion of the gauge fields. The reason is that due to~(\ref{dual_fields}) we exchange
transverse with longitudinal modes in the definitions of $\tilde{A}$ and $\tilde{C}$ and
that the current densities only carry transverse modes. Let us be more precise
the variation of a term $\tilde{A}_\mu X^\mu$ reads
\be
\ba{rcl}
\displaystyle\frac{\delta\tilde{A}_\mu}{\delta A_\nu}\,X^{\mu}&=&\displaystyle\left(\frac{\delta\tilde{F}_{\alpha\beta}}{\delta\tilde{A}_\mu}\right)^{-1}\frac{\delta\tilde{F}_{\alpha\beta}}{\delta F_{\delta\rho}}\,\frac{\delta F_{\delta\rho}}{\delta A_\nu}\,X^\mu\\[5mm]
                                                 &=&\displaystyle 8\epsilon_{\mu\ \,\alpha}^{\ \,\nu\ \,\beta}\left(\partial_\alpha\right)^{-1}\partial_\beta\,X^\mu
\ea
\lb{var_S_dual}
\ee
Now considering the gauge invariance condition (continuity condition) for current densities
$d*J=\partial_\mu J^\mu=0$ we obtain that the currents are given in terms of a regular
anti-symmetric 2-tensor $\phi$ (a 2-form) as
\be
J^\mu=\epsilon^{\mu\delta\rho\lambda}\partial_\delta\phi_{\rho\lambda}+c^\mu
\lb{J_decomp}
\ee
where $c_\mu$ is a constant. This same result is already expressed in~\cite{quant,Berkovits}.
We note that the above expression is obtained from the
Hodge decomposition of the current densities $J=d\varphi+*d\phi+c$. Then replacing
this expression for $X=J$ in the above action variation~(\ref{var_S_dual}) we have the
derivatives in $\mu$ and $\delta$ contracted with the anti-symmetric tensor. Therefore we obtain
a null variation. We note that although we may generally consider non-regular fields, we cannot
consider non-regular current densities, the continuity equation for currents $\partial_\mu J^\mu=0$
is demanded everywhere for gauge invariance, while for the gauge fields $F$ and $G$ are gauge invariant
independently of $A$ and $C$ being regular or not (as long as the gauge transformation parameter is regular,
well understood). The second point to stress is that for regular gauge fields this term is a total derivative,
however for non-regular gauge fields it is not. So by admitting the existence of non-regular gauge
fields the term is present in the action and cannot be integrated to the boundary.

To clarify we give a explicit example.
Let us rewrite the first term of the above expression~(\ref{S_Sources_dual}) in terms of $\phi$ as given
in~(\ref{J_decomp}) as
\be
S_\phi=-\frac{\hat{\epsilon}}{g}\int_MG_{\mu\nu}\phi^{\mu\nu}
\ee
being as usual $G_{\mu\nu}=\partial_\mu C_\nu-\partial_\nu C_\mu$.
One can notice that for regular fields we would integrate by
parts obtaining $S_\phi=-\hat{\epsilon}/g\int_MC_\mu\partial_\nu\phi^{\mu\nu}=0$ because
$\partial_\nu\phi^{\mu\nu}=0$. However take as an example of a non-regular
gauge field $C_1=H[x_2]$ and all the remaining components null, $C_0=C_2=C_3=0$. Here
$H(x)$ is the Heaviside function (also known as unit step function). Then
the above action reads
\be
S_\phi=-\frac{\hat{\epsilon}}{g}\int_M\delta(x_2)\phi^{21}=-\frac{\hat{\epsilon}}{g}\int dtdx^1dx^3dx^4\phi^{21}\neq 0\ .
\ee
Clearly we are not allowed to integrate by parts for non-regular gauge fields.
However when computing the equations of motion for $S_\phi$ we obtain
upon a functional derivation on $C_\mu$ the null contribution for the
equations of motion $\partial_\nu\phi^{\mu\nu}=0$ as desired.

As a last remark we note that adding a current carrying both electric
and magnetic charges (corresponding to a dyon) we obtain an explicit $P$ and $T$ violation
\be
S_{\mathrm{Mix\ Sources}}=-\int_M\left(\frac{1}{e}A_\mu-\frac{1}{g}C_\mu\right)J_{eg}^\mu\ .
\ee
This violation is independent of electromagnetic duality by the simple fact that
$J_{eg}$ must be a combination both of a vector and a pseudo-vector.
So we are assuming that we have no dyons, meaning particles with both electric and
magnetic charge. If they do exist then $P$ and $T$ are not valid symmetries~\cite{Jackson}. 

In the next subsection we derive the Lorentz force checking that we actually have
the usual expression but with the electric and magnetic fields defined as in~(\ref{EB}).

\subsection{Lorentz Force and the Physical Fields}

In order to derive the Lorentz force consider the Lagrangian for a relativistic classical electron with charge $-e$
described by the current density $J_e^\mu=-e(1,\vb{\dot{x}})$
\be
{\mathcal{L}}_{\mathrm{Lorentz-e}}=-m\gamma^{-1}-\left(\frac{1}{e}A_\mu-\frac{\hat{\epsilon}}{g}\tilde{C}_\mu\right)\,J_e^{\mu}
\ee
where the first term accounts for the rest mass and as usual $\gamma^{-1}=\sqrt{1-\dot{x}^2}$. We
have set $c=1$. Varying this action with respect to the coordinates $x_i$ is equivalent to the
Euler-Lagrange equations and we obtain after a straight forward computation that
\be
\ba{rcl}
\displaystyle\frac{dp^i}{dt}&=&\displaystyle+e\left[\left(\frac{1}{e}F^{0i}-\frac{\hat{\epsilon}}{g}\tilde{G}^{0i}\right)+\dot{x}_j\left(\frac{1}{e}F^{ij}-\frac{\hat{\epsilon}}{g}\tilde{G}^{ij}\right)\right]\\[5mm]
               &=&\displaystyle+e\left[E^i+\epsilon^{ijk}\dot{x}_j\,B_k\right]\ .
\ea
\lb{Lorentz-e_Force}
\ee
Where we used the definition of the dual fields $\tilde{G}$ as given in~(\ref{dual_fields}) and
$E^i$ and $B_i$ are given by~(\ref{EB_phys}).

If instead we consider the Lagrangian for a relativistic classical
magnetic monopole with charge $+g$ and current given by $J_g^\mu=+g(1,\vb{\dot{x}})$ we obtain
\be
{\mathcal{L}}_{\mathrm{Lorentz-g}}=-m\gamma^{-1}+\left(\frac{1}{g}C_\mu+\frac{1}{e}\tilde{A}_\mu\right)\,J_g^{\mu}\ .
\ee
Then we obtain
\be
\ba{rcl}
\displaystyle\frac{dp^i}{dt}&=&\displaystyle+g\left[\left(\frac{1}{g}G^{0i}+\frac{1}{g}\tilde{F}^{0i}\right)+\dot{x}_j\left(\frac{1}{g}G^{ij}+\frac{1}{e}\tilde{F}^{ij}\right)\right]\\[5mm]
               &=&\displaystyle+g\left[B^i-\epsilon^{ijk}\dot{x}_j\,E_k\right]\ .
\ea
\lb{Lorentz-g_Force}
\ee
Where again we used the definition of the dual fields $\tilde{F}$ as given in~(\ref{dual_fields}) and
$E^i$ and $B_i$ are given by~(\ref{EB_phys}). We note that here
we considered a positive magnetic charge with rest energy positive, for that reason we obtain a plus sign
in the definition of the Lorentz force.

We note that both Lorentz forces are duality invariant~\cite{Sing_03}.

\ \\
{\large\bf Acknowledgements}

This work was supported by SFRH/BPD/17683/2004. The author thanks the referee
for comments and suggestions.


\begin{thebibliography}{99}
\bibitem{Dirac} P. A. M. Dirac, \textit{Quantized Singularities in the Electromagnetic Field}, Proc. Roy. Soc. {\bf A113} (1931) 60; Phys. Rev. {\bf 74}, \textit{The Theory of Magnetic Poles}, (1948) 817.
\bibitem{CF} N. Cabibbo and E. Ferrari, \textit{Quantum Electrodynamics with Dirac Monopoles}, Il Nuovo Cimento {\bf XXIII} (1962) 1147-1154.
\bibitem{Schwinger} J. Schwinger, \textit{Magnetic Charge and Quantum Field Theory}, Phys. Rev. {\bf 144} (1966) 1087.
\bibitem{Zwa_01} D. Zwanzinger, \textit{Quantum Field Theory of Particles with both Electric and Magnetic Charges}, Phys. Rev. {\bf 176} (1968) 1489-1495.
\bibitem{Zwa_02} D. Zwanzinger, \textit{Local-Langrangian Quantum Field Theory of Electric and Magnetic Charges}, Phys. Rev. {\bf D3} (1971) 880-891.
\bibitem{Zwa_03} R. A. Brandt, F. Neri and D. Zwanzinger, \textit{Lorentz Invariance from Classical particle Paths in Quantum Field Theory of Electric and Magnetic Charges}, Phys. Rev. {\bf D19} (1979) 1153-1167.
\bibitem{Sing_01} D. Singleton, \textit{Does Magnetic Charge Imply a Massive Photon}, Int. J. Theor. Phys. {\bf 35} (1996) 2419-2426, \texttt{hep-th/9509157}.
\bibitem{outro} P. Castelo Ferreira, \textit{Effective Electric and Magnetic Local Actions for Electromagnetism with two Gauge Fields}, \texttt{hep-th/0510078}.
\bibitem{action_00} P. C. R. Cardoso de Mello, S. Carneiro e M. C. Nemes, \textit{Action Principle for the Classical dual Electrodynamics}, Phys. Lett. {\bf B384} 197, \texttt{hep-th/9609218}.

\bibitem{Jackson} J. D. Jackson, \textit{Classical Electrodynamics}, $2^{nd}$ Edition, John Wiley \& Sons, 1975. See section~6.11 and~6.12.


\bibitem{Sing_02} D. Singleton, \textit{Topological Electric Charge}, Int. J. Theor. Phys. {\bf 34} (1995) 2453, \texttt{hep-th/9701040}.
\bibitem{olive} D. I. Olive, \textit{Exact Electromagnetic Duality}, Nucl. Phys. Proc. Suppl. {\bf 45A} (1996) 88-102; Nucl. Phys. Proc. Suppl. {\bf 46} (1996) 1-15, \texttt{hep-th/9508089}


\bibitem{Sing_03} D. Singleton, \textit{Electromagnetism with Magnetic Charge and Two Photons}, Am. J. Phys. {\bf 64} (1996) 452.

\bibitem{new_01} R. L. Packman, \textit{Schwarz-Sen Duality Made Fully Local}, Phys. Lett. {\bf B474} (2000) 309, \texttt{hep-th/9912057}.

\bibitem{PS} Peskin and Schroeder, \textit{Quantum Field Theory},  $1^{st}$ Edition, John Wiley \& Sons, 1997. See section~3.5 for a discussion on spin-statistics and commutation relations. See section~19.1 for axial current anomalies in QED.
\bibitem{Linde} A. D. Linde, \textit{The Universe Multiplication and the Cosmological Constant Problem}, Phys. Lett. {\bf B200} (1988) 272.

\bibitem{Phantom} R. R. Caldwell, \textit{Phantom Menace?}, Phys. Lett. {\bf B545} (2002) 23, \texttt{astro-ph/9908168}.

\bibitem{Tito_01} P. Castelo Ferreira and J. Tito Mendon\c{c}a, \textit{Generalized Proca Equations and Vacuum Current from Breaking of $U(1)\times U(1)$ Gauge Symmetry}, \texttt{hep-th/0601171}.
\bibitem{Tito_02} J. Tito Mendon\c{c}a and P. Castelo Ferreira, \textit{Mass for Plasma Photons from Gauge Symmetry Breaking}, Europhys. Lett. {\bf 75} (2006) 189, \texttt{hep-th/0601166}.

\bibitem{YW} T. T. Wu and C. N. Yang, \textit{Concept of Nonintegrable Phase Factors and Global Formulation of Gauge Fields}, Phys. Rev. {\bf D12} (1975) 3845; \textit{Dirac's Monopole Without Strings: Classical Lagrangian Theory}, Phys. Rev. {\bf D14} (1976) 437

\bibitem{Sing_04} A. Kato and D. Singleton, \textit{Gauging Dual Symmetry}, Int. J. Phys. {\bf 41} (2002) 1563, \texttt{hep-th/0106277}.

\bibitem{Witten} E. Witten, \textit{Dyons of Charge $e\theta/2\pi$}, Phys. Lett. {\bf B86} (1979) 283-287.


\bibitem{action_01} R. W. Kuhne, \textit{A Model of Magnetic Monopoles}, Mod. Phys. Lett. {\bf A12} 3153, \texttt{hep-th/9708394}.
\bibitem{action_02} C. A. P. Galv\~ao, J. A. Mignaco, \textit{A Consistent Electromagnetic Duality}, \texttt{hep-th/0002182}.



\bibitem{Berkovits} N. Berkovits, \textit{Local Actions with Electric and Magnetic Sources}, Phys. Lett.  {\bf B395} (1997) 28-35, \texttt{hep-th/9610134}.
\bibitem{quant} S. Carneiro, \textit{Probability Amplitudes for Charge-Monopole Scattering}, JHEP 9807 (1998) 022, \texttt{hep-th/9702036}.




\end{thebibliography}
\end{document}